\documentclass{aa}
\usepackage{epsfig}

\newcommand{\magpt}[2]{\mbox{$\rm #1\hspace{-0.25em}\stackrel{m}{.}
      \hspace{-1.0mm}#2$}}                             

\def\bsec{\hbox{$.\!\!{\arcsec}$}}
\def\deg{\hbox{$.\!\!{^\circ}$}}

\newcommand{\Rsun}{\mbox{$R\sb{\odot}$}}

\newcommand\teff{$ {\rm T_{eff}}$}
\newcommand\Teff{$ {\rm T_{eff}}$}

\newcommand\logg{$\log {\rm g}$}

\def\gtrsim{\mathrel{\hbox{\rlap{\hbox{\lower4pt\hbox{$\sim$}}}\hbox{$>$}}}}
\def\lesssim{\mathrel{\hbox{\rlap{\hbox{\lower4pt\hbox{$\sim$}}}\hbox{$<$}}}}

\newcommand{\Msolar}{\mbox{\,$\rm M_{\odot}$}}        
\begin{document}
\title{Spectral Types and Masses of White Dwarfs in
Globular Clusters\thanks{Based on observations collected at the
European Southern Observatory, Chile (ESO proposal 65.H-0531, 67.D-0201)}
\thanks{Based on observations with the NASA/ESA
{\it Hubble Space Telescope} obtained at the Space Telescope Science
Institute, which is operated by the Association of Universities for
Research in Astronomy, Inc., under NASA contract NAS 5-26555}}
 \author{S. Moehler\inst{1,4}
 \and D. Koester\inst{1}
 \and M.~Zoccali\inst{2,5} \and F.R. Ferraro\inst{3} \and U.~Heber\inst{4}
 \and R.~Napiwotzki\inst{4} \and A. Renzini\inst{5}}
\offprints{S. Moehler}
\institute{
Institut f\"ur Theoretische Physik und Astrophysik der Universit\"at Kiel, 
24098 Kiel, Germany
(e-mail: moehler,koester@astrophysik.uni-kiel.de)
\and Departamento de Astronom\'{\i}a y Astrof\'{\i}sica, Pontificia 
 Universidad Cat\'olica de Chile, Avenida Vicu\~na Mackenna 4860,
 782-0436 Macul, Santiago, Chile (e-mail: mzoccali@astro.puc.cl)
\and Dipartimento di Astronomia, Universit\`a di Bologna, via Ranzani
1, 40127 Bologna, Italy (ferraro@bo.astro.it)
\and Dr. Remeis-Sternwarte, Astronomisches Institut der Universit\"at
Erlangen-N\"urnberg, Sternwartstr. 7, 96049 Bamberg, Germany
(heber@sternwarte.uni-erlangen.de,rn38@astro.le.ac.uk)
\and
European Southern Observatory, Karl Schwarzschild Strasse 2, 85748
Garching bei M\"unchen, Germany (arenzini@eso.org)
}
\titlerunning{White Dwarfs in Globular Clusters}
\authorrunning{Moehler et al.}
\date{Received 8 December 2003, accepted 2 March 2004}

\abstract{White dwarfs in globular clusters offer additional
possibilities to determine distances and ages of globular clusters,
provided their spectral types and masses are known. We therefore
started a project to obtain spectra of white dwarfs in the globular
clusters NGC~6397 and NGC~6752.  All observed white dwarfs show
hydrogen-rich spectra and are therefore classified as DA. Analysing
the multi-colour photometry of the white dwarfs in NGC~6752 yields an
average gravity of \logg\ = 7.84 and 0.53\Msolar\ as the most probable
average mass for globular cluster white dwarfs. Using this average
gravity we try to determine independent temperatures by fitting the
white dwarf spectra.  While the stellar parameters determined from
spectroscopy and photometry usually agree within the mutual error
bars, the low resolution and S/N of the spectra prevent us from
setting constraints stronger than the ones derived from the photometry
alone. For the same reasons the white dwarf spectra obtained for
NGC~6397 unfortunately do not provide an independent distance estimate
of sufficient accuracy to distinguish between the long and short
distance  scale for globular clusters.
\keywords{Stars: white dwarfs -- globular clusters: individual:
NGC~6397 -- globular clusters: individual: NGC~6752}} \maketitle

\section{Introduction} 
As white dwarfs are the final stage in the evolution of all low mass
stars, many are expected to exist in globular clusters.  However, due
to their faintness and occurrence in very crowded fields they were
detected only after the refurbishment of HST with the WFPC2. Several
candidate white dwarfs were then soon identified in M15 (de Marchi \&
Paresce \cite{depa95}), $\omega$ Cen (Elson et al. \cite{elgi95}), NGC
6397 (Paresce et al. \cite{pade95a}, Cool et al. \cite{copi96}), M4
(Richter et al. \cite{rifa95}, \cite{ribr02}), 47 Tuc (Paresce et
al. \cite{pade95b}, Zoccali et al. \cite{zore01}) and NGC 6752
(Renzini et al. \cite{rebr96}). Renzini \& Fusi Pecci (\cite{refu88})
suggested to use the white dwarf sequence as a standard candle for
determining the distance to nearby globular clusters, similarly to the
traditional main sequence fitting procedure using local subdwarfs with
known trigonometric parallax. In this case, the white dwarf sequence
of the cluster is compared to a sequence constructed with local white
dwarfs with accurate trigonometric parallax. The method was then
applied to NGC 6752 (Renzini et al. \cite{rebr96}) and 47 Tuc (Zoccali
et al. \cite{zore01}, where also the result from Renzini et
al. \cite{rebr96} for NGC~6752 was updated). However, while the
updated distance modulus for NGC~6752 agrees well with all other
distance determinations for this cluster, the white dwarf distance to
47~Tuc is considerably shorter than that found by Gratton et
al. (\cite{grfu01}, \cite{grbr03}) via main sequence fitting.

 While it may seem strange to use the faintest objects in a
globular cluster to derive its distance white dwarfs offer
some advantages as standard candles when compared to main
sequence stars:
\begin{itemize}
\item They come in just two varieties - either hydrogen-rich (DA) or
helium-rich (DB) -- {\em independent of their original metallicity}
 and, in both cases, their atmospheres are
virtually free of metals. So, unlike in the case
of main sequence fitting, there is not the problem of finding local
calibrators with the same metallicity as the globular clusters.
\item White dwarfs are locally much more numerous than metal-poor main
sequence stars and thus allow to define a better reference sample.
\end{itemize}

However, the method has its own specific problems, which are discussed
in great detail in Zoccali et al. (\cite{zore01}) and Salaris et
al. (\cite{saca01}). Indeed, the location of the white dwarf cooling sequence
depends on:

\begin{itemize}
\item {\em the white dwarf mass}\\ 
On theoretical grounds, given the observed
maximum luminosity reached on the asymptotic giant branch (AGB), the
mass of currently forming white dwarfs in globular clusters should be
$0.53\pm 0.02$\Msolar\ (Renzini \& Fusi Pecci \cite{refu88}, Renzini et
al. \cite{rebr96}). Unfortunately, there are no local white
dwarfs in this mass range with directly determined masses
(i.e. without using a mass-radius relationship).
There is, however, a handful of local white dwarfs with
spectroscopically determined masses near this value (cf. Table 1 in
Zoccali et al. \cite{zore01}), which allows to construct a semi-empirical
cooling sequence for $M_{\rm WD}$=0.53\Msolar, once relatively small
mass-dependent corrections are applied to each local white dwarf.
The spectroscopic determination of the mass of white dwarfs in a
globular cluster was first attempted by Moehler et al. (\cite{mohe00})
for white dwarfs in NGC~6397. However, the low S/N of the spectra of
these very faint stars did not allow to determine the mass with
sufficient accuracy.

\item {\em the white dwarf envelope mass}\\ In the case of DA
white dwarfs the cooling sequence location depends also on
the mass of the residual hydrogen-rich envelope. This affects 
{\it spectroscopically} derived masses (see above),
with the resulting mass being $\approx 0.04$\Msolar\ higher when using
the {\it evolutionary} envelope mass ($\approx 10^{-4}$M$_{\rm WD}$,
Fontaine \& Wesemael \cite{fowe97}) as
opposed to virtually zero envelope mass. This mass uncertainty
corresponds to an uncertainty of \magpt{0}{1} in the distance modulus
and 1--1.5 Gyr in the age derived from the main sequence turnoff.

\item {\em spectral type}\\ DB stars are fainter than DA stars at a
given colour, with the offset depending on the filter combination
(i.e. the offset is greater in $V$ vs. $B-V$ than in $I$
vs. $V-I$). However, more massive DA white dwarfs are also fainter at
a given colour.
\end{itemize}

The white dwarf sequence also allows to determine the age of a
globular cluster if its faint end is detected. In that case one can derive
the age of the globular cluster from the luminosity of its oldest and
faintest white dwarfs. Aside from the observational difficulties and
the uncertainties in the cooling tracks (see Chabrier et
al. \cite{chbr00} for more details) any error in the assumed
mass affects the result. Recent very deep HST observations allowed to
detect the white dwarf cooling sequence in M~4 to unprecedented depths
of $V$$\approx$30 (Richer et al. \cite{ribr02}).  As a preliminary
result Hansen et al. (\cite{habr02}) derive an age of 12.7$\pm$0.7 Gyr
from the white dwarf luminosity function of M~4, consistent with other
independent age estimates (their error bar, however, does not include
errors due to the uncertainty of the white dwarf mass). Their result
has recently been questioned by de Marchi et al. (\cite{depa03}), who
claim that the cluster membership of the white dwarfs can not be
verified down to sufficiently faint limits to obtain more than a lower
limit of the age. Richer et al. (\cite{ribr04}) showed, however,
that different methods of data reduction and analysis account for
the different depths reached with the same data and thereby defended
the original result of Hansen et al. (\cite{habr02}).

In view of the relevance of the white dwarf masses and spectral
types to the problems described above we decided to observe spectra
of white dwarfs in NGC~6397 and NGC~6752 in order to determine their
spectral types and get mass estimates. Pilot observations of the white
dwarf candidates in NGC~6397 (63.H-0348) had already shown that the
targets are hydrogen-rich DA white dwarfs (Moehler et
al. \cite{mohe00}), but did not allow much quantitative work.

\section{Observations and Data Reduction} 
\subsection{Photometry\label{sec-phot}}
The dataset used for NGC~6752 in the present work consist of a
series of HST/WFPC2 exposures taken through the filters F336W, F439W,
F555W, F814W in March and April 1995 as a part of the HST programme
GO5439 (see Table~\ref{tab-HST}). Results based on this dataset have
been already published in Renzini et al. (\cite{rebr96}) and Ferraro
et al. (\cite{feca97}). The field is centered about 2\arcmin\
south-east of the cluster center (see Fig.~1 in Ferraro et
al. \cite{feca97}).

\begin{table}
\caption[]{Description of the used photometric data-sets.
The filter name, the number of exposures and the total integration
time (in seconds) are given. \label{tab-HST}}
\begin{tabular}{lrr}
\hline
\hline
Filter & No of exp & Exp. time\\
\hline
  F336W &  12 &    26,400 \\
  F439W &  5 &    10,000 \\
  F555W &   5 &    6,000  \\
  F814W &   7 &    7,000  \\
\hline
\end{tabular}
\end{table}

All the photometric reductions have been carried out using ROMAFOT
(Buonanno et al. \cite{bubu83}), a package specifically developed to
perform accurate photometry in crowded fields.  A version of ROMAFOT
optimized to handle sub-sampled HST images has been used for the
present work (see Buonanno \& Iannicola \cite{buia88}).  Details of
the reduction procedure can be found in Ferraro et
al. (\cite{feca97}).  In short, we combined all the images in each
filter obtaining a median image free of cosmic ray events and other
defects.  In order to perform a deep search for faint blue objects we
used the combined F439W image (with an equivalent total exposure time
of 10,000 sec) as the reference frame for object detection. The object
list derived from the median F439W image was used as input for the
photometry of the median images in the other filters and preliminary
colour-magnitude diagrams (CMDs) have been obtained.  The preliminary
CMDs have been used to select a list of objects that fell outside the
main CMD sequences. A visual inspection revealed that most of them
were spurious objects (e.g., artifacts due to the diffraction patterns
of bright saturated stars, hot pixels, etc). By using the {\it
cleaned} object list, the standard PSF fitting procedure was finally
performed on each single image by using a Moffat (\cite{moff69})
function plus a numerical map of residuals in order to better account
for the contribution of the stellar PSF wings.

A final catalogue was then compiled with coordinates and instrumental
magnitudes for each star in each filter. The magnitudes were
calibrated to the HST flight system following Dolphin (\cite{dolp00}).
Specifically, we applied: 
\begin{enumerate}
\item The charge transfer efficiency (CTE) correction, which is
negligible for the brightest white dwarfs, but reaches up to
\magpt{0}{25} for the faintest ones in the F336W filter 
\item The aperture corrections, computed by means of a sample of
isolated stars in each WFPC2 chip and filter, in order to insure that
the instrumental magnitudes refer to the counts in a 0\bsec5 radius
around each star
\item The gain correction, due to the fact that the four WFPC2 chips
do not have identical response factors 
\item A correction to the $U$ magnitude, removing the effect of the decreasing
UV sensitivity of the CCD 
\item The flight system zero points, given in Table~6 of Dolphin
(\cite{dolp00}) 
\end{enumerate}

\begin{figure}
\epsfxsize=0.5\textwidth
\epsffile{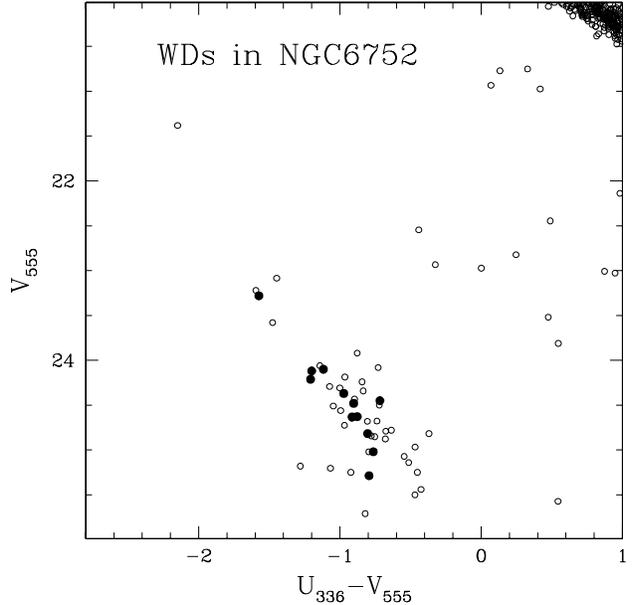}
\caption[]{$V$ vs $U-V$ (Vega-based 
flight system) for the stars in NGC~6752. The
stars listed in Table~\ref{HST-phot} targets are marked by filled
circles.\label{cmd}}
\end{figure}

The resulting colour-magnitude diagram is shown in
Fig.~\ref{cmd} and clearly shows the white dwarf sequence. The
photometric data for the twelve most isolated white dwarfs in
NGC~6752, from which the spectroscopic targets were selected,  are
listed in Table~\ref{HST-phot}.

\begin{table*}
\caption[]{The photometric data for the twelve most isolated white dwarfs
in NGC~6752 in the Vega-based HST flight system\label{HST-phot}}
\begin{tabular}{lllllllll}
\hline
\hline
Star & $U_{336}$& $\Delta U_{336}$ & $B_{439}$ & $\Delta B_{439}$ &
$V_{555}$ & $\Delta V_{555}$ & $I_{814}$ & $\Delta I_{814}$ \\
\hline
WF2-241  & \magpt{24}{014} & \magpt{0}{040} & \magpt{25}{052} & \magpt{0}{020}
 & \magpt{24}{818} & \magpt{0}{030} & \magpt{24}{916} & \magpt{0}{060}\\
WF2-3648 & \magpt{22}{919} & \magpt{0}{020} & \magpt{24}{152} & \magpt{0}{020}
 & \magpt{24}{119} & \magpt{0}{030} & \magpt{24}{195} & \magpt{0}{040}\\
WF2-5639 & \magpt{21}{707} & \magpt{0}{010} & \magpt{23}{261} & \magpt{0}{020}
 & \magpt{23}{281} & \magpt{0}{020} & \magpt{23}{447} & \magpt{0}{030} \\
WF3-296  & \magpt{24}{493} & \magpt{0}{080} & \magpt{25}{419} & \magpt{0}{020}
 & \magpt{25}{287} & \magpt{0}{040} & \magpt{24}{770} & \magpt{0}{070}\\
WF3-1610 & \magpt{23}{004} & \magpt{0}{030} & \magpt{24}{321} & \magpt{0}{020}
 & \magpt{24}{212} & \magpt{0}{020} & \magpt{24}{127} & \magpt{0}{030}\\
WF3-3584 & \magpt{23}{718} & \magpt{0}{040} & \magpt{24}{908} & \magpt{0}{020}
 & \magpt{24}{632} & \magpt{0}{020} & \magpt{24}{639} & \magpt{0}{040} \\
WF3-4909 & \magpt{24}{256} & \magpt{0}{070} & \magpt{25}{295} & \magpt{0}{020}
 & \magpt{25}{020} & \magpt{0}{040} & \magpt{24}{923} & \magpt{0}{050}\\
WF3-6441 & \magpt{23}{577} & \magpt{0}{040} & \magpt{24}{574} & \magpt{0}{020}
 & \magpt{24}{480} & \magpt{0}{030} & \magpt{24}{613} & \magpt{0}{040} \\
WF3-6454 & \magpt{23}{750} & \magpt{0}{050} & \magpt{24}{895} & \magpt{0}{020}
 & \magpt{24}{628} & \magpt{0}{040} & \magpt{24}{266} & \magpt{0}{060} \\
WF4-1700 & \magpt{22}{983} & \magpt{0}{030} & \magpt{24}{210} & \magpt{0}{020}
 & \magpt{24}{100} & \magpt{0}{020} & \magpt{24}{067} & \magpt{0}{040}\\
WF4-2003 & \magpt{23}{733} & \magpt{0}{040} & \magpt{24}{730} & \magpt{0}{020}
 & \magpt{24}{450} & \magpt{0}{030} & \magpt{24}{577} & \magpt{0}{030}\\
WF4-2081 & \magpt{23}{397} & \magpt{0}{040} & \magpt{24}{520} & \magpt{0}{020}
 & \magpt{24}{370} & \magpt{0}{020} & \magpt{24}{522} & \magpt{0}{040}\\
\hline
\end{tabular}
\end{table*}

\subsection{Spectroscopy\label{sec_spec}}
The spectroscopic targets in NGC~6397 (65.H-0531(B)) were selected from the
photometry by Cool et al. (\cite{copi96}), those in NGC~6752
(67.D-0201(B)) from the
data discussed in Sect.~\ref{sec-phot}. We took care to select stars
as isolated as possible in order to avoid contamination
from neighbouring stars. The spectra were obtained in service mode with
FORS1 at VLT-UT1 (Antu) and VLT-UT3 (Melipal) on the dates given in
Table~\ref{tab-obs}.  We used the MOS mode of FORS1 (slit length
19\arcsec) with the high resolution collimator (0\bsec1/pixel), a slit
width of 0\bsec8 and grism V300. The detector was a TK2048EB4-1
backside thinned CCD with 2048$\times$2048 pixels of (24 $\mu$m)$^2$,
which was read out with high gain (1.46 e$^-$/count, 5.15 e$^-$
read-out noise, NGC~6397) respectively low gain (2.87 e$^-$/count,
6.07 e$^-$ read-out noise, NGC~6752) and normal read-out speed using
one read-out port only.  This configuration yields a reciprocal dispersion of
2.5~\AA/pixel and a wavelength coverage of at least 3500~\AA\ to
5700~\AA\ (depending on the position of the slitlet also larger
coverage is possible).

The slit width is larger than the requested seeing of 0\bsec6 to allow
for small positioning inaccuracies, but results in a seeing-dependent
resolution in those cases where the actual seeing is smaller than the
slit width (cf. Table~\ref{tab-obs}). As discussed below, due to the
low resolution and S/N of our spectra the change in resolution due to varying
seeing does not affect the results of the spectral fitting. As FORS1 is
equipped with an atmospheric dispersion corrector MOS observations at
higher airmass are not a problem.

The spectra were supposed to be observed only if the seeing was better
than 0\bsec6 (corresponding to a resolution of 15~\AA). However, due
to the faintness of our targets we had to use rather long exposure
times (2955 sec for NGC~6397 and 4380 sec for NGC~6752), during which
the seeing might have changed. A few exposures were stopped before reaching
the requested exposure time due to technical problems.  The values
provided in Table~\ref{tab-obs} are averaged over the exposure time
unless noted otherwise. Some spectra were taken at too bad seeing
($\ge$1\arcsec) or with rather bright and nearby moon. Those
observations provided no useful data and are therefore not listed in
Table~\ref{tab-obs}.

\begin{table*}
\caption[]{Observational parameters for the spectroscopic data. On
August 10, 2001, the seeing monitor did not work and the seeing was
estimated from the acquisition and through-slit images
\label{tab-obs}}
\begin{tabular}{lllllrl}
\hline
\hline
target & start of observation & exp. time & 
seeing & airmass & \multicolumn{2}{c}{moon}\\
&  & & [sec]  &  & distance & illumination \\
\hline
NGC~6397 & 2000/07/01 03:18:24.011 & 2955 & 0\bsec53 & 1.149 & 
 147\deg4 & 0.007 \\
NGC~6397 & 2000/07/01 04:09:25.965 & 2955 & 0\bsec54 & 1.148 &
 147\deg4 & 0.006 \\
NGC~6397 & 2000/07/05 03:44:29.402 & 2955 & 0\bsec72 & 1.146 &
 115\deg0 & 0.153 \\
NGC~6397 & 2000/07/05 04:35:30.895 & 2955 & 0\bsec87 & 1.174 &
 114\deg5 & 0.156 \\
NGC~6397 & 2000/07/06 05:42:39.878 & 2955 & 0\bsec80 & 1.282 &
 102\deg2 & 0.254 \\
NGC~6397 & 2000/07/06 06:33:39.752 & 2955 & 0\bsec81 & 1.425 &
 101\deg7 & 0.259 \\
NGC~6397 & 2000/07/08 03:49:31.331 & 2955 & 0\bsec56 & 1.151 &
  80\deg5 & 0.453$^1$\\
NGC~6397 & 2000/07/08 04:40:30.861 & 2955 & 0\bsec50 & 1.192 &
  80\deg1 & 0.457$^1$\\
NGC~6397 & 2000/07/25 02:52:02.328 & 2955 & 0\bsec53 & 1.156 &
 124\deg1 & 0.422$^2$\\
NGC~6397 & 2000/07/25 03:43:02.144 & 2955 & 0\bsec53 & 1.204 &
 124\deg5 & 0.417$^2$\\
\hline
NGC~6752 & 2001/08/09 01:16:57.397 & 1756 & 0\bsec67  & 1.260 &
  86\deg1 & 0.796$^3$\\
NGC~6752 & 2001/08/09 01:53:59.110 & 1724 & 0\bsec61  & 1.235 &
  86\deg4 & 0.794$^3$\\
NGC~6752 & 2001/08/10 00:57:05.415 & 4380 & 0\bsec63  & 1.263 &
  95\deg6 & 0.715$^4$\\
NGC~6752 & 2001/08/10 02:16:43.582 & 2400 & 0\bsec67  & 1.229 &
  96\deg2 & 0.710$^4$\\
NGC~6752 & 2001/08/11 00:30:47.431 & 4380 & 0\bsec65: & 1.287 &
 105\deg0 & 0.626$^5$\\
NGC~6752 & 2001/08/11 01:50:27.188 & 4380 & 0\bsec55: & 1.235 &
 105\deg7 & 0.620$^5$\\
NGC~6752 & 2001/08/11 03:18:27.263 & 4380 & 0\bsec50: & 1.280 &
 106\deg3 & 0.613$^5$\\
NGC~6752 & 2001/08/11 04:35:06.749 & 4380 & --        & 1.418 &
 106\deg9 & 0.608$^5$\\
NGC~6752 & 2001/08/22 00:53:16.938 & 4380 & 0\bsec70  & 1.237 &
  94\deg6 & 0.119\\
NGC~6752 & 2001/08/22 02:16:26.569 & 4380 & 0\bsec56  & 1.262 &
  93\deg9 & 0.124\\
NGC~6752 & 2001/09/15 00:04:58.643 & 4380 & 0\bsec64  & 1.239 &
 132\deg8 & 0.092\\
NGC~6752 & 2001/09/15 01:37:23.372 & 4380 & 0\bsec60  & 1.332 &
 132\deg1 & 0.086\\
NGC~6752 & 2001/09/18 00:25:56.746 & 4380 & 0\bsec65  & 1.257 &
 100\deg0 & 0.006\\
NGC~6752 & 2001/09/18 01:41:59.797 & 4380 & 0\bsec53  & 1.363 &
  99\deg3 & 0.007\\
NGC~6752 & 2001/09/19 00:11:09.587 & 4380 & 0\bsec70  & 1.250 &
  88\deg2 & 0.035\\
\hline
\multicolumn{6}{l}{$^1$ moon rise at 11:58 UT}\\
\multicolumn{6}{l}{$^2$ moon rise at 00:47 UT}\\
\multicolumn{6}{l}{$^3$ moon rise at 02:05 UT}\\
\multicolumn{6}{l}{$^4$ moon rise at 02:56 UT}\\
\multicolumn{6}{l}{$^5$ moon rise at 03:49 UT}\\
\end{tabular}
\end{table*}

For NGC~6752 dome flat fields with two different illumination patterns
were observed for each night. For the previous programme on NGC~6397,
however, we had dome flat fields (again with two different
illumination patterns) only for the nights of July 3, 8, and 24, 2000.
For data from the other nights we used the flat field closest in time.
As part of the standard calibration we were also provided with
masterbias frames for our data. The masterbias showed no evidence for
hot pixels and was smoothed with a 30$\times$30 box filter to keep any
possible large scale variations, but erase noise. The flat fields were
averaged for each night and bias-corrected by subtracting the smoothed
masterbias of the night. From the flat fields we determined the limits
of the slitlets in spatial direction. Each slitlet was extracted and
from there on treated like a long-slit spectrum. The flat fields were
normalized along the dispersion axis with a linear fit, which
corrected most of the large scale variations.

Due to the low resolution of the data we obtained a wavelength
calibration spectrum with a slit width of 0\bsec3 (the center
positions of the slitlets were unchanged) to allow a better definition
of the line positions. The dispersion relation for all spectra was
obtained from this one calibration spectrum by fitting 3$^{\rm rd}$ or
4$^{\rm th}$ order polynomials along the dispersion axis. We used 13
to 16 unblended lines between 3600~\AA\ and 6200~\AA\ and achieved an
r.m.s. error of about 0.1~\AA\ per CCD row.

Due to the long exposure times the scientific observations contained a
large number of cosmic ray hits. Those were corrected with the
algorithm described in G\"ossl \& Riffeser (\cite{gori02}), using a
threshold of 15$\sigma$ and a FWHM of 1.5 pixels for the cosmic ray
hits. As the routine is not intended originally for the use with
spectra we also reduced the uncleaned frames to allow a check for any
possible artifacts of the cosmic cleaning procedure, but found none. 
The slitlets
with the stellar spectra were extracted in the same way as the flat
field and wavelength calibration slitlets. The smoothed masterbias was
subtracted and they were divided by the corresponding normalized flat
fields, before they were rebinned to constant wavelength steps. 

\begin{figure}
\epsfxsize=0.5\textwidth
\epsffile{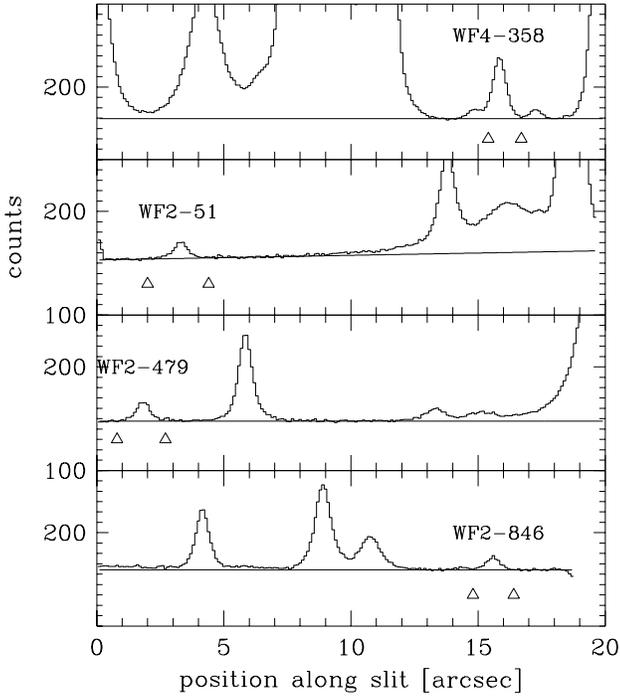}
\caption[]{Spatial brightness profile of the stars observed in
NGC~6397 for each slitlet. The histogram profiles were obtained by
averaging the two-dimensional spectra between 4000~\AA\ and 5000~\AA. The
straight lines mark the profile of the fitted sky background averaged
over the same wavelength range. The triangles mark the spatial range
over which the white dwarf spectra were extracted.\label{slit6397}}
\end{figure}

\begin{figure}
\epsfxsize=0.5\textwidth
\epsffile{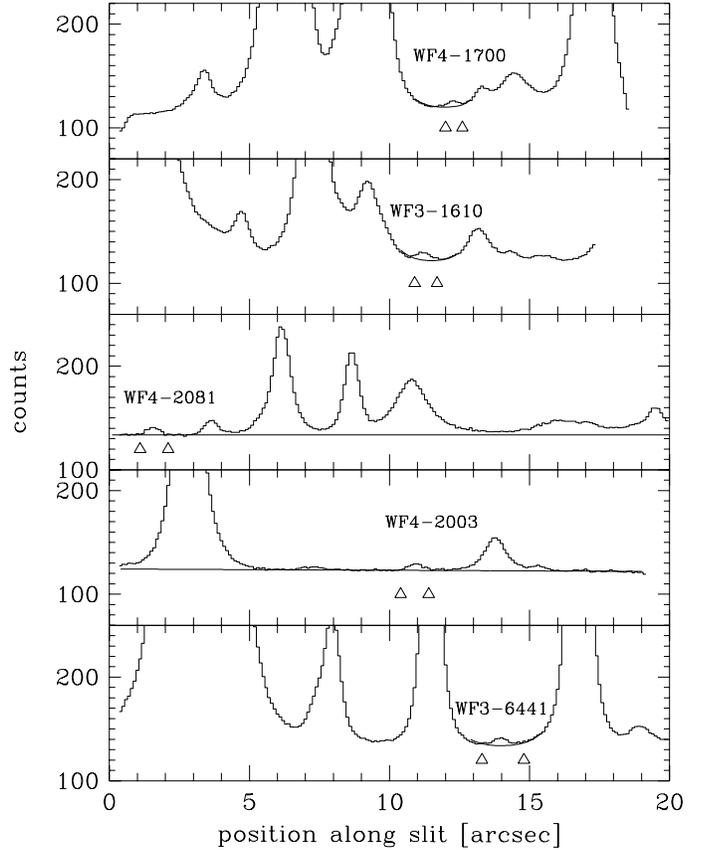}
\caption[]{Spatial brightness profile of the stars observed in
NGC~6752 for each slitlet. The profiles were obtained by averaging the
two-dimensional spectra (histograms) respectively the two-dimensional
fitted background (smooth lines) between 4000~\AA\ and 5000~\AA. The
slitlets containing WF3-6441, WF3-1610, and WF4-1700 clearly show the
effect of light from neighbouring stars providing a spatially varying
background at the white dwarf's position. The triangles mark the
spatial range over which the white dwarf spectra were
extracted.\label{slit6752}}
\end{figure}

For the NGC6397 data we then filtered the rebinned frames along the
spatial axis with a median filter of 7 pixels (corresponding to
0\bsec7) width to erase any remaining cosmic ray hits and reduce the
noise in the sky background. We then identified regions uncontaminated
by any stellar source in the filtered data (where the profiles of
the stars are of course widened) and approximated the spatial
distribution of the sky background by a constant or a linear fit
(cf. Fig.~\ref{slit6397}). The sky background spectra obtained
this way from the filtered data were then subtracted from the {\em
unfiltered} two-dimensional spectra.

In the case of NGC~6752 the observed regions were in three cases
(WF4-1700, WF3-1610, and WF3-6441) so crowded that the wings of
spectra of brighter stars contributed to the background of the white
dwarf spectra (cf. Fig.~\ref{slit6752}).  To account for this varying
background we fitted the wings of the brighter stars with Lorentzian
profiles. The FWHM and position of the two Lorentzian profiles were
determined from averaged spatial profiles like those shown in
Fig.~\ref{slit6752} and the maximum was allowed to vary with
wavelength to account for the spectral features of the bright
stars. For WF4-2081 and WF4-2003 we performed the sky background
subtraction in the same way as described for NGC~6397.

For both clusters the sky-subtracted spectra were extracted using
Horne's (\cite{horn86}) algorithm as implemented in MIDAS (Munich
Image Data Analysis System).  Finally the spectra were corrected for
atmospheric extinction using the extinction coefficients for La Silla
(T\"ug \cite{tueg77}) as implemented in MIDAS. As can be seen from
Table~\ref{tab-obs} we observed a large number of spectra for each
target, which had to be co-added. Before averaging the spectra for
each star we had to ensure that they are all on the same wavelength
scale.  As the wavelength calibration spectra are obtained during
daytime slight shifts between the slitlet positions of scientific and
calibration observations may occur. Due to their very low S/N 
we cannot check for offsets using the individual
white dwarf spectra themselves. To account for possible zero-point
shifts in wavelength we cross-correlated the range 5500~\AA--5650~\AA\
in the sky background spectra of all scientific observations with the
first spectra obtained on the night of June 30, 2000 (NGC~6397) and
August 13, 2001 (NGC~6752).  If the mean shift for all spectra from
one exposure was more than 0.5~\AA\ (corresponding to 0.2 pixels) we
corrected all spectra from that exposure with the mean offset. This
correction however, cannot account for small offsets of the stars'
positions from the center of the slit in dispersion direction. The
signal-to-noise ratio of the individual spectra was much too low to
obtain these offsets via radial velocities at a resolution of about
15~\AA.  For the stars in NGC~6752 we were lucky to have a blue
horizontal branch star in one of the slitlets. We used the spectrum of
this star to determine any remaining velocity shifts and used
those to correct the spectra of the white dwarfs to laboratory
wavelengths. In NGC~6397 we used the radial velocity offset derived
from the spectrum of WF4-358 for each observation to correct all other
spectra.

\begin{figure}
\epsfxsize=0.5\textwidth
\epsffile{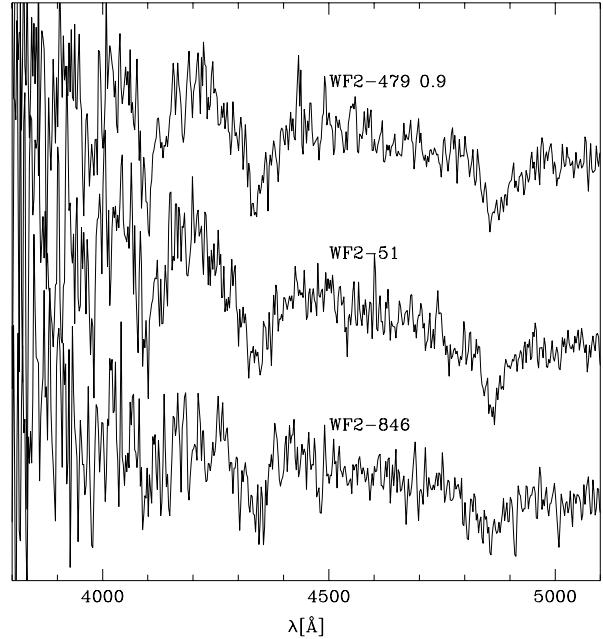}
\caption[]{Spectra of the three fainter white dwarfs in 
NGC~6397\label{spec6397}}
\end{figure}

\begin{figure}
\vspace*{6.8cm}
\includegraphics{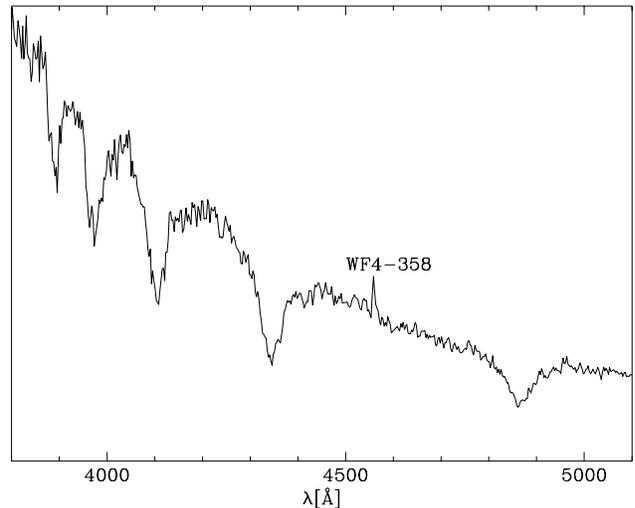}
\caption[]{Spectrum of the brightest white dwarf observed 
in NGC~6397\label{spec4_358}}
\end{figure}

\begin{figure}
\epsfxsize=0.5\textwidth
\epsffile{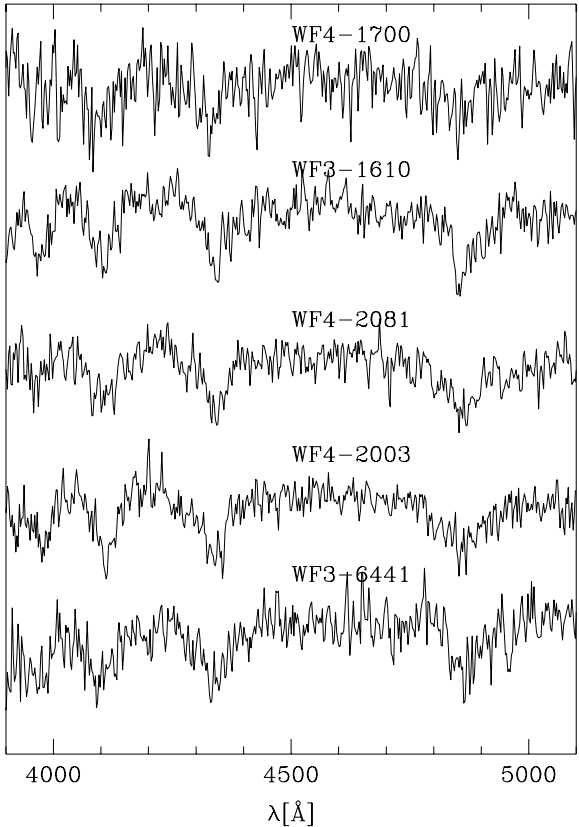}
\caption[]{Spectra of the white dwarfs in NGC~6752 (normalized)
\label{spec6752}}
\end{figure}

For a relative flux calibration we used response curves derived from
spectra of LTT~6248, LTT~7987, and LTT~9239 with the data of Hamuy et
al. (\cite{hawa92}) for NGC~6752. For NGC~6397 we used the standard
stars LTT~7379, Feige~110 (Hamuy et al. \cite{hawa92}) and GD~248 (Oke
\cite{oke90}). The response curves were fitted by splines and averaged
for all nights. The white dwarf spectra of NGC~6752 were normalized
with a curve fitted to the continuum of the blue HB star spectrum,
which had also been used for the determination of the velocity shifts.
The resulting spectra are shown in Figs.~\ref{spec6397},
\ref{spec4_358} and \ref{spec6752}, with the brightness of the stars
decreasing from top to bottom. The region below 3800~\AA\ is
completely dominated by noise, so that neither a spectral slope nor
spectral lines can be identified. This may be explained by the low UV
throughput of FORS1, the reasons for which are not understood.  Also
the brightest star, WF4-358, however, shows no spectral lines below
3800~\AA.

\section{Fit of the photometry of NGC6752}

\subsection{Theoretical magnitudes for the HST Photo\-metry}

To calculate theoretical magnitudes in the so called ``STMAG''
system of the WFPC2 (Holtzman et al. \cite{hobu95}) we used the grid
of DA model atmospheres of Koester (see Finley et al. \cite{fiko97}
for a detailed description) and the following definition of $M_{\rm
STMAG}$

$$
M_{\rm STMAG} = -2.5\, 
{\frac{\int S(\lambda) F_{\lambda}\,d\lambda}{\int S(\lambda)d\lambda}} - 21.1
$$
where $S(\lambda)$ is the filter response function and
$F_{\lambda}$ is the mean intensity of the stellar disk or $ 4
H_{\lambda}$ in terms of the Eddington flux. Note that $M_{\rm STMAG}$ is not
an absolute magnitude but is used to distinguish theoretical
from observed magnitudes $m$. With this definition of STMAG the
magnitudes are all zero ($M_{\rm STMAG} = 0)$ for a star with a constant
$F_{\lambda}$ of $3.63 \,10^{-9}$ erg\,s$^{-1}$\,cm$^{-2}$\,\AA$^{-1}$. 

A star observed in this system should have magnitude $m_{\rm STMAG} = 0$,
if the flux arriving at the earth is $f_{\lambda} = 3.63\,10^{-9}$ in
the same units. The relation between observed and theoretical
magnitude can be obtained from
\begin{eqnarray*}
        f_{\lambda} &  = & \frac{\pi R^2}{d^2}\, F_{\lambda} \\
  m_{\rm STMAG} &=& -2.5 \log\frac{\pi R^2}{d^2} + M_{\rm STMAG} +21.1 \\   
            &=& M_{STMAG} + (m - M) - 5\,\log\frac{R}{\Rsun} + 41.991
\end{eqnarray*} 
Here, $R$ is the radius of the star, $d$ the distance, and $(m - M)$
the apparent distance modulus for the respective filter band. 

The observed magnitudes for the white dwarfs in the globular clusters
were not determined in the STMAG system, but rather in the so called
``WFPC2 Flight System'' (Holtzman et al. \cite{hobu95}; Dolphin
\cite{dolp00}). We have chosen to transform the observations to the
STMAG system by applying the difference between the two zeropoints
(STMAG from Holtzman et al. \cite{hobu95} and FS from Dolphin
\cite{dolp00}) before the fitting. The applied corrections (STMAG-FS)
are \magpt{0}{128}, \magpt{-0}{724}, \magpt{-0}{041}, \magpt{1}{228}
for the magnitudes $U_{336}, B_{439}, V_{555}, I_{814}$
respectively. We have tested this procedure in the following way:

First we calculated the expected magnitudes for Vega in the STMAG
system, using the absolutely calibrated Vega spectrophotometry from
the HST archive (Colina et al. \cite{cobo96}). The inverse of the
above corrections were then applied to the Vega STMAG magnitudes,
which should bring them on the Flight System, with the resulting
values \magpt{-0}{008}, \magpt{0}{076}, \magpt{0}{039}, \magpt{0}{002}
for $U_{336}, B_{439}, V_{555}, I_{814}$. The Flight System is defined
in such a way that it should match the magnitudes of the ground based
system used for pre-flight calibrations for stars with zero colors
(see Holtzman et al. \cite{hobu95} for a detailed discussion). Vega
has approximately zero colors, but the magnitudes -- though coming
close -- do not exactly match the observed $UBVI$ magnitudes, which
Holtzman et al. (\cite{hobu95}) take as \magpt{0}{02}, \magpt{0}{02},
\magpt{0}{03}, \magpt{0}{035}. One probable reason for this difference
is that the calibration between flight system and ground system was
built using stars (mainly white dwarfs) with a spectral type very
different from that of Vega. The former, in fact, have much higher
surface gravities and stronger Balmer lines.

Although it would be possible to make additional corrections in an
attempt to bring the system of the globular cluster white dwarfs
closer to the Vega system, we do not believe that this is worthwhile
in view of several other calibration uncertainties and the observational
errors for the faint objects.

\subsection{Fitting observations with theoretical magnitudes}
\begin{figure}
\epsfxsize=0.5\textwidth
\epsffile{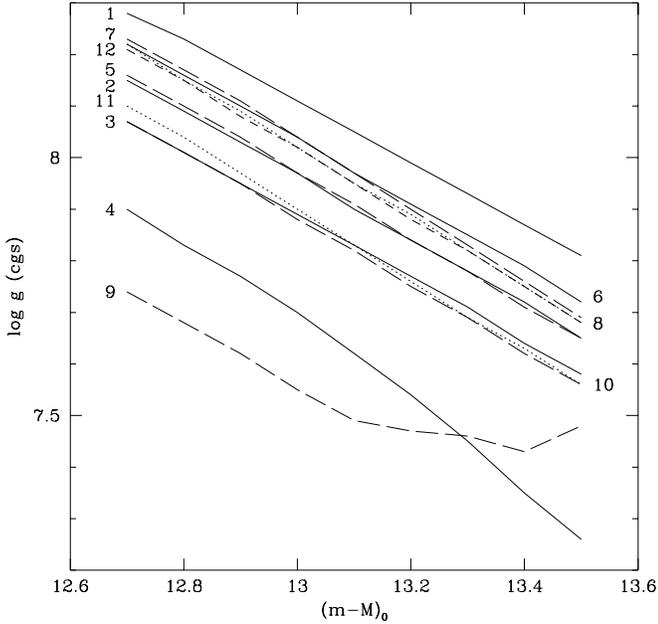}
\caption[]{Surface gravities determined from the photometric data
listed in Table~\ref{par-phot} for various distance moduli. The
numbers refer to the position of the star in Table~\ref{par-phot}
\label{wd_phot_logg}}
\end{figure}

When trying to fit a set of observed magnitudes with theoretical
values we have in principle the possibility to determine
simultaneously the three parameters \Teff, \logg\ and
distance. Effective temperature and surface gravity determine $M_{\rm
STMAG}$ as described above and provide the radius $R$ of the white
dwarf via the Wood (\cite{wood95}) mass-radius relation. In practice
it turned out to be impossible to determine all three free parameters
from only four observed magnitudes. In addition to observational
errors, the major problem is the strong degeneracy with respect to the
parameters. The {\em relative} energy distribution for the white
dwarfs in the observed parameter range depends mostly on \teff\
and only very little on \logg. The {\em solid angle} ($\pi R^2/d^2$),
which determines the difference between observed and theoretical
magnitudes, depends strongly on both \logg\ (through the mass-radius
relation) and on the distance. Therefore a small change in \logg\ can
be easily compensated by a change in distance and vice versa.
Fig.~\ref{wd_phot_logg} shows the strong correlation between derived surface
gravity and distance modulus.

It was therefore necessary, to keep one or two parameters fixed at
some pre-determined value, to achieve a robust fitting result. The
fitting used a $\chi^2$ routine based on the Levenberg-Marquardt
algorithm (Press et al. \cite{prte92}) and is described in more detail in
Zuckerman et al. (\cite{zuko03}). Before the fit, the magnitudes were
dereddened assuming E$_{\rm B-V}$ = \magpt{0}{04}, and the extinction
coefficients of Holtzman et al. (\cite{hobu95}).

\begin{table*}
\caption[]{Fitting the photometry of the twelve most isolated
white dwarfs in NGC6752 with
\teff\ and \logg\ as free parameters, (m-M)$_0$ fixed at \magpt{13}{05} and
\magpt{13}{20}. E$_{\rm B-V}$ assumed as \magpt{0}{04}.
\label{par-phot}}
\begin{tabular}{lrrrr|rrrr}
\hline
\hline
 Object & \multicolumn{4}{c|}{(m-M)$_0$ = \magpt{13}{05}}
& \multicolumn{4}{c}{(m-M)$_0$ = \magpt{13}{20}}\\
 &\teff& $\Delta$\teff &\logg & $\Delta$\logg& 
 \teff& $\Delta$\teff &\logg & $\Delta$\logg \\
\hline
WF2-241  &  13000 & 600 &   8.08 & 0.07  &  13100 & 500 &   7.99 & 0.06 \\
WF2-3648 &  17400 & 500 &   7.94 & 0.04  &  17500 & 400 &   7.84 & 0.04 \\
WF2-5639 &  25200 & 500 &   7.86 & 0.03  &  25200 & 500 &   7.77 & 0.03 \\
WF3-296  &   9400 &1500 &   7.66 & 0.46  &   9400 &1500 &   7.54 & 0.49 \\
WF3-1610 &  16700 &1700 &   7.94 & 0.14  &  16700 &1700 &   7.84 & 0.15 \\
WF3-3584 &  13400 &1000 &   8.00 & 0.11  &  13600 &1100 &   7.91 & 0.12 \\
WF3-4909 &  11300 & 200 &   8.00 & 0.02  &  11200 & 300 &   7.90 & 0.05 \\
WF3-6441 &  14700 &1300 &   7.99 & 0.14  &  14700 &1200 &   7.89 & 0.12 \\
WF3-6454 &  10400 &1900 &   7.52 & 0.45  &  10700 &2000 &   7.47 & 0.51 \\
WF4-1700 &  16200 & 900 &   7.85 & 0.08  &  16300 &1000 &   7.75 & 0.09 \\
WF4-2003 &  13000 & 800 &   7.86 & 0.09  &  13100 & 700 &   7.76 & 0.09 \\
WF4-2081 &  15300 & 700 &   7.98 & 0.06  &  15400 & 600 &   7.88 & 0.05 \\
\hline
weighted mean & &  &   7.96 & 0.02 & 
      &  &   7.84 & 0.02 \\
\hline
\end{tabular}
\end{table*}

We decided to keep the distance modulus fixed as it is rather well
known for NGC~6752: The distance modulus obtained by Renzini et
al. (\cite{rebr96}) using white dwarfs in the cluster was
\magpt{13}{05}. A reevaluation of these white dwarf observations,
using the ``thick hydrogen envelope'' models results in \magpt{13}{15}
(Zoccali et al. \cite{zore01}). Recent determinations with main
sequence fitting obtain \magpt{13}{17} (Reid \cite{reid98}) and
\magpt{13}{23} (Gratton et al. \cite{grfu01}). Gratton et
al. (\cite{grbr03}) recently published a new distance determinations
for NGC~6752 from main sequence fitting, which yielded (m-M)$_0$ =
\magpt{13}{23}\ from Johnson photometry and (m-M)$_0$ = \magpt{12}{99}
from Str\"omgren photometry. As it is unclear from their data which
value is the correct one we do not use this new determination for our
average distance. For our fits we have used the short distance
originally determined by Renzini et al. (\cite{rebr96},
\magpt{13}{05}) as an extreme case, 
as well as the average of the more recent
determinations not involving white dwarfs (\magpt{13}{20}).

The results listed in Table~\ref{par-phot} show that the scatter in
\logg\ is quite small due to the very strong correlation between
the radius (\logg) and distance through the solid angle of the star
(see above). The average value for the surface gravity is 7.96 for
(m-M)$_0$ = \magpt{13}{05} and 7.84 for (m-M)$_0$ = \magpt{13}{20}
with an error of the mean of 0.02. With the Wood (\cite{wood95})
mass-radius relation for a typical \teff\ of 15000~K and ``thick
hydrogen layer'' ($10^{-4}$ of the stellar mass) this corresponds to
masses of 0.59\Msolar\ and 0.53\Msolar, for the short and long
distance modulus, respectively. Assuming instead the more unlikely
case of a ``thin'' hydrogen layer ($\le 10^{-6}$ M$_{\rm WD}$) the
mass would be 0.56\Msolar\ for (m-M)$_0$ = \magpt{13}{05} and
0.50\Msolar\ for (m-M)$_0$ = \magpt{13}{20}.  If we consider that the
progenitor stars in globular clusters should have less than 1\Msolar\
and that NGC~6752 has an exclusively blue horizontal branch with a
very extended blue tail, the smaller mass seems much more likely and
gives strong support for the larger of the two distances, in
agreement with most distance determinations for NGC~6752. Using the
distance modulus as variable parameter and \logg\ as fixed (7.96,
7.84) yields essentially the same results, as expected.  The effective
temperature of the objects does not change significantly with the
various methods, and we may conclude that it is fairly well
constrained.

\section{Spectral fits for NGC~6752 and NGC~6397} 
The observed spectra for the white dwarfs in the two clusters were
analysed with the same model atmosphere grid used for the calculation
of the theoretical magnitudes. In order to verify the influence of
different fit programs on the results (and thus get at least some
estimate of the minimum true errors as opposed to the purely
statistical errors provided by the fit routines) we used the procedure
described in Finley et al. (\cite{fiko97}, marked by ``DK'') and that
described in Napiwotzki et al. (\cite{nagr99}, marked by ``RN''). 

The low S/N and resolution of the spectra 
leads to very large errors if \teff\ and
\logg\ are both used as free parameters. We have therefore decided to
keep \logg\ fixed at the value 7.84 obtained as the most likely value
from the photometry of NGC~6752. For the spectral fits we used a
resolution of 15~\AA, corresponding to a seeing of 0\bsec6 (which is
the average seeing of our observations). 

\begin{figure}[!h]
\epsfxsize=0.45\textwidth
\epsffile{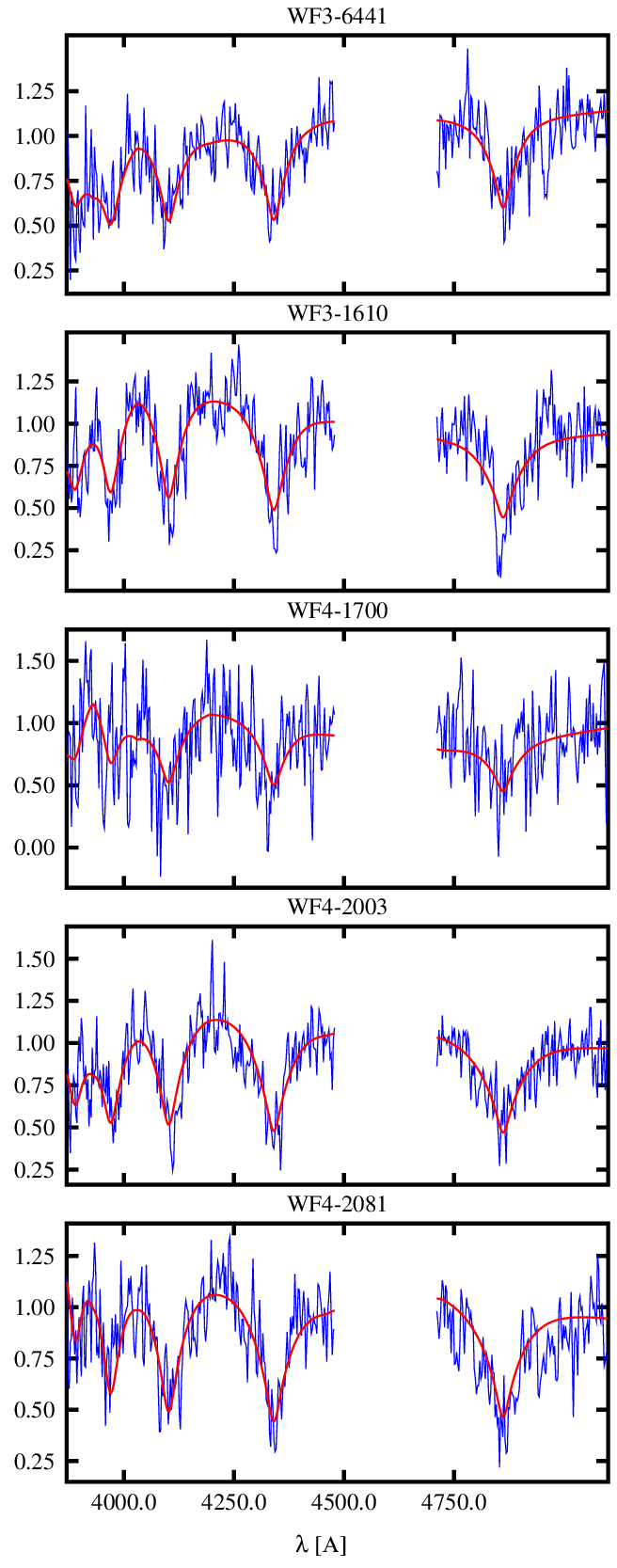}
\caption[]{Spectral fits of the white dwarfs in
NGC~6752. Only the part of the spectrum, which was
actually fitted, is shown.\label{fitspec1}}
\end{figure}

\begin{figure*}[!ht]
\epsfxsize=0.9\textwidth
\epsffile{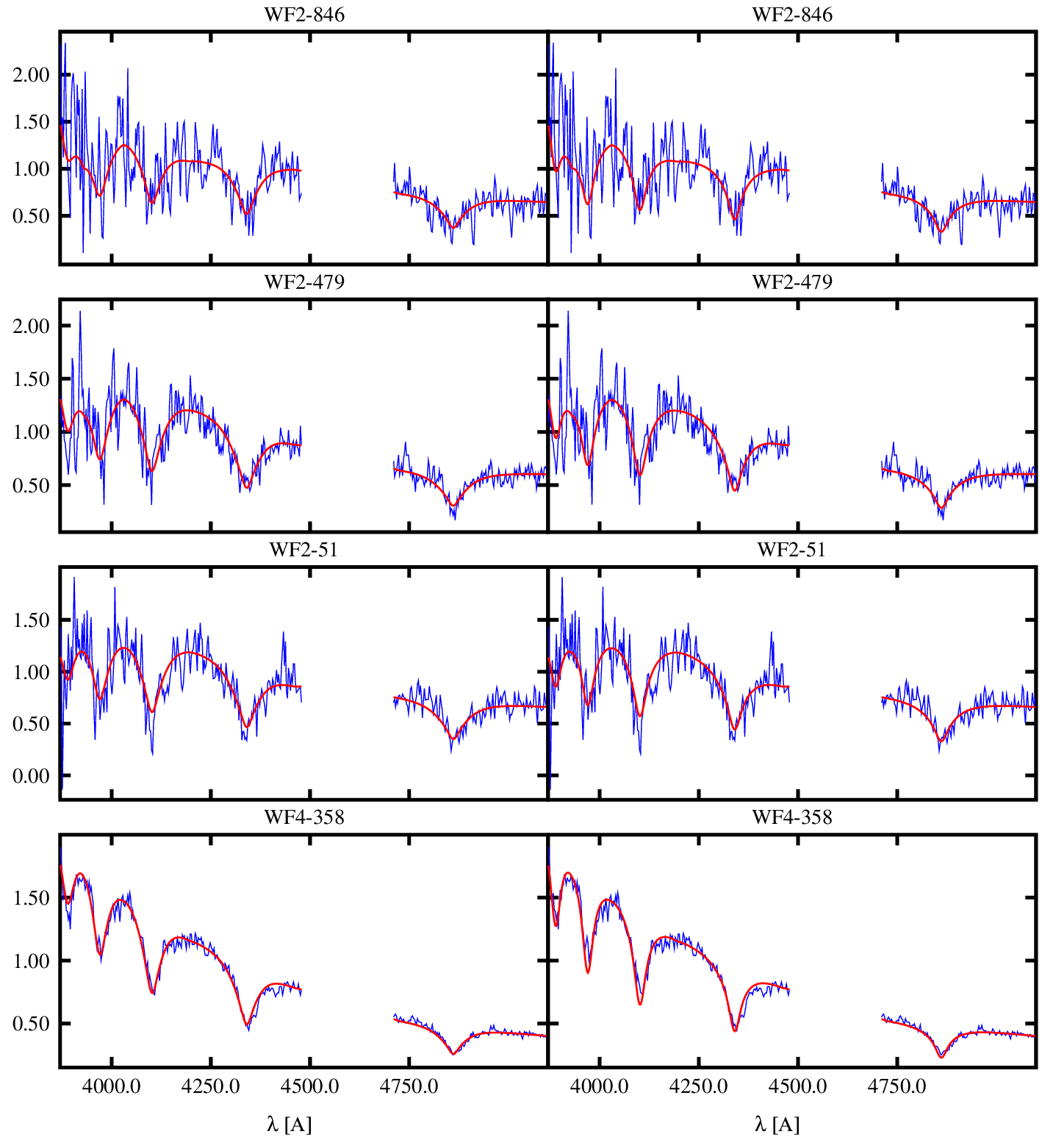}
\caption[]{Spectral fits of the white dwarfs in NGC~6397.
Only the part of the spectrum, which was
actually fitted, is shown. The left panels show the hot solutions, the 
right panels show the cool solutions.\label{fitspec3}}
\end{figure*}

To allow a better understanding of our results and their errors
we discuss here some technical details of the fitting procedures as
applied in this case. When searching for the best-fitting model (by
the $\chi^2$ fitting routine) in a first step the model is fitted to
the observations at pre-selected ``continuum'' points. In the DK
fitting routine these are 7 points located in the continuum between H$_\gamma$
and H$_\beta$ and longward of H$_\beta$ in the red part of the
spectrum, and in the middle between Balmer lines in the blue part. The
observed continuum is determined from the median of the flux in a
region 20 -- 100 \AA\ wide around these points. For each of these
continuum wavelengths a normalizing factor is calculated, which would
adjust the model flux to the observations at these points. The
complete normalized model spectrum is then obtained by interpolating
these normalization factors. Since this uses a linear interpolation the
normalized model can show abrupt changes in slope from one interval to
the next if the observation is badly calibrated or extremely noisy, as
is apparent at the blue ends of some of the spectra in
Figs.~\ref{fitspec1} and \ref{fitspec3}.  After this normalization the
$\chi^2$ is determined from the comparison of model and observations
in several predetermined intervals, which normally include the
spectral line profiles. In the present case we have used the two
intervals from 3870 -- 4480 and 4710 -- 5100 \AA. These are the
intervals plotted in the figures; the pure continuum region between
4480 and 4710 \AA\ does not contain any useful information since the
normalization always forces agreement.

The major difference between the DK and the RN method is the
definition of the
continuum points. RN used two continuum regions of
40\,\AA\ width adjacent to the fitted regions of the H$_\beta$, H$_\gamma$,
and H$_\delta$ lines. As in the DK-case normalization factors between
model and observed flux are computed and linearly interpolated for
each individual Balmer line. Since the high noise level combined with
the strong overlapping of the lines makes the definition of continuum
points between lines more and more difficult for the higher members of
the Balmer series, the normalization of the region shortward of and
including H$_\epsilon$ was done for all lines simultaneously. Four continuum
regions (of 30\,\AA\ width) were defined from the red wing of
H$_\epsilon$ to 3815~\AA. 
A linear fit of the resulting normalization factors was
performed and applied to the fitted lines.

\begin{table*}
\caption[]{Fitting the white dwarf spectra in NGC~6752
with \logg\ held fixed at 7.84. ${\rm T_{eff,ph}}$ is the result from the
photometry for (m-M)$_0$ = \magpt{13}{20}.
\label{par-spec1}}
\begin{tabular}{llrlrlr}
\hline
\hline
 Object & \multicolumn{2}{c}{DK} & \multicolumn{2}{c}{RN} & & \\
 & \teff & $\Delta$\teff & \teff & $\Delta$\teff & ${\rm T_{eff,ph}}$
& $\Delta {\rm T_{eff,ph}} $\\
 & [K] & [K] & [K] & [K] & [K] & [K] \\
\hline
WF3-1610 & 15100 & 900 & 14400$^1$ & 1000 & 16700 & 1700\\
WF3-6441 & 19200  & 1000 & 21600 & 1300 & 14700 & 1200\\
WF4-1700 & 18500 & 1600 & 16800 & 1800 & 16300 & 1000\\
WF4-2003 & 13400 & 1700 & 13400 & 2700 & 13100 & 700\\
WF4-2081 & 12900 & 800 & 14600 & 800 & 15400 & 600\\
\hline 
\end{tabular}\\
\begin{tabular}{l}
$^1$ excluding the core of H$_\beta$\\
\end{tabular}
\end{table*}

Finally the best-fitting model is in both cases found by minimizing
$\chi^2$ as a function of the model parameters $\log$~g and
\Teff. This routine automatically determines errors for the fit
parameters, which are however only statistical errors, assuming that
the differences between the model and the observation are caused only
by statistical measurement errors distributed normally around the
correct value. There are, however, a number of sources of systematic
errors, which cannot be determined easily. These include the reduction
procedures, with errors in our case likely being dominated by the
background subtraction. It is obvious that for an observation with a
background both variable and several times higher than the continuum
spectrum a small error in the background determination can have a profound
influence on the line profile and thus the fitting results. Placing
the continuum points differently, or interpolating the factors
quadratically instead of linearly will also change the results, but
without prior knowledge of the exact observed spectrum there is no
reason to prefer one method over the other. Such systematic errors can
only be estimated by comparing results from different spectra for the
same star, or comparing results from different authors and fitting/reduction
methods.

To verify the influence of the assumed resolution on the results we
also performed fits for resolution of 10\AA\ and 20\AA, which yielded
effective temperatures hotter respectively cooler by less than 1\%,
which is well below even the purely statistical errors given in
Table~\ref{par-spec1}.  The observed spectra (confined to the ranges
used for the model fits) and the best fitting models for NGC~6752 are
shown in Fig.~\ref{fitspec1}.

The effective temperatures derived for the white dwarfs in
NGC6752 with different fit methods, although slightly different, usually
agree within the errorbars (cf. Table~\ref{par-spec1}) and we
therefore conclude that the choice of the fit methods does not affect our
results.  For WF3-6441 the spectroscopic temperatures 
agree with each other but differ from the photometric one. 
This may be an effect of the rather problematic sky subtraction
for this star (cf. Fig.~\ref{slit6752} and
Sect.~\ref{sec_spec}). Despite our best efforts to approximate the
spatial and spectral variation of the sky background in these crowded
fields, we cannot exclude the possibility of residual background light 
in the extracted spectra. As the flux in the spectra is significantly
lower than that of the background small errors in the background
approximation can strongly affect the spectra. We therefore rather
trust the photometric results for this star, despite its larger than
average error.

\begin{table*}
\caption[]{Fitting the white dwarf spectra in NGC~6397 
with \logg\ held fixed at 7.84. ${\rm T_{eff,ph}}$ is the effective
temperature estimated from $VI$ photometry by Moehler et
al. (\cite{mohe00}). We also give M$_V$ from the model fit (assuming a
thick hydrogen layer), the observed $V$ corrected for a reddening of
E$_{\rm B-V}$ = \magpt{0}{18}, and the derived distance modulus.
\label{par-spec2}}
\begin{tabular}{lrllrrrlrrr}
\hline
\hline
 Object & V$_0$ & ${\rm T_{eff,ph}}$ &
\multicolumn{4}{c}{DK} & \multicolumn{4}{c}{RN} \\
& & & \teff & $\Delta$\teff & M$_V$ & (m-M)$_0$ & \teff & $\Delta$\teff & 
M$_V$ & (m-M)$_0$ \\
 & & [K] & [K] & [K] & & & [K] & [K] & & \\
\hline
WF2-846  & \magpt{23}{76} & 10300 & 10700 & 400 & \magpt{11}{73} &
\magpt{12}{03} & 11400 & 700 & \magpt{11}{56} & \magpt{12}{20}\\
 & & & 17900 & 1200 & \magpt{10}{73} & \magpt{13}{03} & 
15700 & 1000 & \magpt{10}{96} & \magpt{12}{80}\\
WF2-479  & \magpt{23}{33} & 11500 & 11500 & 400 & \magpt{11}{54} &
\magpt{11}{79} & 12700 & 800 & \magpt{11}{35} & \magpt{11}{98} \\ 
& & & 15300 & 600 & \magpt{11}{01} & \magpt{12}{32} & 
14700 & 800 & \magpt{11}{07} & \magpt{12}{26}\\
WF2-51   & \magpt{23}{46} & 10800 & 11200 & 400 & \magpt{11}{60} &
\magpt{11}{86} & 12800 & 1000 & \magpt{11}{34} & \magpt{12}{12} \\
 & & & 15500 & 500 & \magpt{10}{98} & \magpt{12}{48} &
12800 & 1000 & \magpt{11}{34} & \magpt{12}{12} \\
WF4-358  & \magpt{22}{19} & 19400 & 10200 & 100 & \magpt{11}{88} &
\magpt{10}{31} & 10600 & 100 & \magpt{11}{76} & \magpt{10}{43} \\
 & & & 19000 & 300 & \magpt{10}{62} & \magpt{11}{57} &
17900 & 300 & \magpt{10}{73} & \magpt{11}{46} \\
\hline
\end{tabular}
\end{table*}

The observed spectra and best fitting models for the white dwarfs in
NGC~6397 are shown in Fig.~\ref{fitspec3}. For these stars we
have only $V$ and $I$ photometry from Cool et al. (\cite{copi96}),
from which we estimated temperatures in Moehler et al. (\cite{mohe00}, 
cf. Table~\ref{par-spec2}). These temperature estimates are very
useful to distinguish between the hot and cool solutions obtained from 
the spectra alone (see Table~\ref{par-spec2}). Based on the photometric
temperature estimates we select the cool results of the spectroscopic fits for
the three fainter stars and the hot result for the brightest star. As
can be seen from Fig.~\ref{fitspec3} the fitted model spectra for the 
hot (left panel) and cool (right panel) solutions are
indistinguishable by eye (except perhaps for the brightest star) and
also the $\chi^2$ values for both solutions are almost identical. As
can be seen from Table~\ref{par-spec2} the cool solutions of the RN
fit are hotter than those from the DK fit, whereas the situation is
just the opposite for the hot solutions, indicating the presence of
systematic errors in our analysis.

The parameters obtained from the spectral fit can be used to determine
absolute magnitudes from the models, and -- with the help of observed
$V$ values -- also individual distances (see Table~\ref{par-spec2}).
The average distance modulus for the three fainter objects is
\magpt{11}{9}$\pm$\magpt{0}{1} [DK] and \magpt{12}{1}$\pm$\magpt{0}{1}
[RN], respectively (errors are statistical r.m.s errors only), yielding a mean 
distance modulus of \magpt{12}{0}, which is at the short end of the
range of distance moduli found in the recent literature
(\magpt{11}{99} -- \magpt{12}{24}).

The last object -- although
the brightest -- is rather discrepant, and may be a foreground object or
have a significantly smaller \logg\ than the others. 
In view of the fact that the proper motions derived 
by King et al. (\cite{kian98}) support the cluster membership of all
four white dwarfs analysed in NGC~6397 we decided to check the gravity
for this star through a fit with both
\teff\ and \logg\ as free parameters (see Table~\ref{par-spec3}).
 Pilot
spectra of this star with a combined exposure time of 1.5 hours
yielded an effective temperature of 18,200$\pm$1300~K and a surface
gravity of \logg\ = 7.3$\pm$0.36 (Moehler et al. \cite{mohe00}). While
the temperatures from all three fits agree very well, the scatter in
surface gravity again shows that the systematic errors discussed
above are
considerably larger than the statistical ones. The lower gravity
obtained by the DK fit would support the cluster membership of the
brightest white dwarf, but would yield a mass of 0.48~\Msolar\ instead 
of the 0.53~\Msolar\ derived for the fainter white dwarfs in
NGC~6752. 

\section{Conclusions}

We observed a sample of white dwarfs in the globular clusters NGC~6397
and NGC~6752 and showed that they are all hydrogen-rich DA. From
multicolour photometry we determined an average mass of 0.53$\pm$0.03\Msolar\ 
(uncertainty due to uncertainties of 0.02 dex
in the average \logg\ and of \magpt{0}{05} in the distance modulus of
NGC~6752). This value (based on the assumption of a thick hydrogen layer)
is identical to that assumed by Renzini et al. (\cite{rebr96})
on theoretical grounds. Therefore both observational and theoretical
arguments strongly advocate against the use of 0.6~\Msolar\ (the mean
mass of the local white dwarfs) or even more for the mass of hot white
dwarfs in globular clusters when comparing observations to theoretical
tracks. However, the limited S/N in combination with the low
resolution of the spectroscopic data prevented the independent
determination of masses from spectroscopic fits alone.  Multi-colour
photometry may be the better way to determine the physical parameters
of white dwarfs in globular clusters once their spectral types are
known. For spectroscopic observations
 our experience shows that crowding
can severely limit the usefulness of the data due to problems with sky
subtraction. We would therefore strongly recommend to look for white
dwarf candidates in ground-based wide-field photometry to avoid the
problems we encountered for NGC~6752. Also a better sensitivity in the 
blue would be useful for future spectroscopy.

\begin{table}[!h]
\caption[]{Results for the brightest object in NGC~6397 
with \logg\ as free parameter. We also give M$_V$ from the model fit
(assuming a thick hydrogen layer), the observed $V$ corrected for a
reddening of E$_{\rm B-V}$ = \magpt{0}{18}, and the derived distance
modulus.
\label{par-spec3}}
\begin{tabular}{lrrrrrr}
\hline
\hline
& \teff & $\Delta$\teff & \logg & $\Delta$ \logg & M$_V$ & (m-M)$_0$  \\
 & [K] & [K] & & & &  \\
\hline
DK & 18800 & 340 & 7.72 & 0.06 & \magpt{10}{47} & \magpt{11}{72}\\
RN & 18700 & 520 & 8.09 & 0.12 & \magpt{11}{02} & \magpt{11}{17}\\
\hline
\end{tabular}
\end{table}

\acknowledgements We want to thank the staff at Paranal observatory
for their great effort in performing these demanding
observations. While in Bamberg S.M. was supported by the BMBF grant
No. 50 or 9602 9. Thanks go also to our referee, Dr. M. van Kerkwijk,
for very valuable suggestions, which improved this paper considerably.

\end{document}